# A note on proper conformal vector fields in cylinderically symmetric static space-times

Ghulam Shabbir

Faculty of Engineering Sciences

GIK Institute of Engineering Sciences and Technology

Topi Swabi, NWFP, Pakistan

Email: shabbir@giki.edu.pk

and

Shaukat Iqbal

Faculty of Computer Science and Engineering

GIK Institute of Engineering Sciences and Technology

Topi Swabi, NWFP, Pakistan

**Abstract**

A study of proper conformal vector field in non conformally flat cylindrically symmetric static space-times is given by using direct integration technique. Using the above mentioned technique we have shown that a very special class of the above space-time admits proper conformal vector field (CVF).

**Keywords**: Direct integration technique; Conformal vector field

## 1. INTRODUCTION

The aim of this paper is to find the existance of conformal vector fields in the non conformally flat cylindrically symmetric static space-times. The conformal vector field which preserves the metric structure upto a conformal factor carries significant interest in Einstein's theory of general relativity. It is therefore important to study this symmetry. Different approaches [1, 3-8] were adopted to study conformal vector fields. In this paper a direct integration technique is used to study conformal vector fields in the non conformally flat cylindrically symmetric static space-times. Throughout $M$ represents a four dimensional, connected, hausdorff space-time manifold with Lorentz metric $g$ of signature (-, +, +, +). The curvature tensor associated with $g_{ab}$, through the Levi-Civita connection, is denoted in component form by $R^a{}_{bcd}$, the Weyl tensor components are $C^a{}_{bcd}$, and the Ricci tensor components are $R_{ab} = R^c{}_{acb}$. The usual covariant, partial and Lie derivatives are denoted by a semicolon, a comma and the symbol $L$, respectively. Round and square brackets denote the usual symmetrization and skew-symmetrization, respectively. The space-time $M$ will be assumed non conformally flat in the sense that the Weyl tensor does not vanish over any non empty open subset of $M$.

Any vector field $X$ on $M$ can be decomposed as

$$X_{a;b} = \frac{1}{2}h_{ab} + F_{ab} \qquad (1)$$

where $h_{ab}(=h_{ba}) = L_X g_{ab}$ and $F_{ab}(=-F_{ba})$ are symmetric and skew symmetric tensors on $M$, respectively. Such a vector field $X$ is called conformal vector field if the local diffeomorphisms $\psi_t$ (for appropriate $t$) associated with $X$ preserve the metric structure up to a conformal factor i.e. $\psi_t^* g = \phi g$, where $\phi$ is a nowhere zero positive function on some open subset of $M$ and $\psi_t^*$ is a pullback map on some open subset of $M$ [3]. This is equivalent to the condition that

$$h_{ab} = \phi g_{ab},$$

equivalently

$$g_{ab,c}X^c + g_{cb}X^c_{,a} + g_{ac}X^c_{,b} = \phi g_{ab}, \qquad (2)$$

where $\phi : U \to R$ is the smooth conformal function on some subset of $M$, then $X$ is a called conformal vector field. If $\phi$ is constant on $M$, $X$ is homothetic (proper homothetic if $\phi \neq 0$) while if $\phi = 0$ it is Killing [1]. If the vector field $X$ is not homothetic then it is called proper conformal. It follows from [3] that for a conformal vector field $X$, the bivector $F$ and the function $\phi$ satisfy (putting $\phi_a = \phi_{,a}$)

$$F_{ab;c} = R_{abcd}X^d - 2\phi_{[a}g_{b]c}, \tag{3}$$

$$\phi_{a;b} = -\frac{1}{2}L_{ab;c}X^c - \phi L_{ab} + R_{c(a}F_{b)}{}^c, \tag{4}$$

where $L_{ab} = R_{ab} - \frac{1}{6}R g_{ab}$.

## 2. Main Results

Consider a cylindrically symmetric static space-time in usual coordinate system $(t,r,\theta,z)$ with line element [2]

$$ds^2 = -e^{V(r)}dt^2 + dr^2 + e^{U(r)}d\theta^2 + e^{W(r)}dz^2. \tag{5}$$

The possible Segre type of the above space-time is {1,111} or one of its degeneracies. The above space-time (5) admits three linearly independent killing vector fields, which are

$$\frac{\partial}{\partial t}, \frac{\partial}{\partial \theta}, \frac{\partial}{\partial z}. \tag{6}$$

A vector field $X$ is said to be a conformal vector field if it satisfies equation (2). One can write (2) explicitly using (5) we have

$$V'(r)X^1 + 2X^0_{,0} = \phi \tag{7}$$

$$X^1_{,0} - e^{V(r)}X^0_{,1} = 0 \tag{8}$$

$$e^{U(r)}X^2_{,0} - e^{V(r)}X^0_{,2} = 0 \tag{9}$$

$$e^{W(r)}X^3_{,0} - e^{V(r)}X^0_{,3} = 0 \tag{10}$$

$$2X^1_{,1} = \phi \tag{11}$$

$$e^{U(r)}X^2_{,1} + X^1_{,2} = 0 \tag{12}$$

$$e^{W(r)}X^3_{,1} + X^1_{,3} = 0 \tag{13}$$

$$U'(r)X^1 + 2X^2_{,2} = \phi \tag{14}$$

$$e^{W(r)}X^3_{,2} + e^{U(r)}X^2_{,3} = 0 \tag{15}$$

$$W'(r)X^1 + 2X^3_{,3} = \phi, \qquad (16)$$

where $\phi = \phi(t,r,\theta,z)$. Equations (11), (8), (12) and (13) give

$$\begin{aligned}
X^0 &= \int e^{-V(r)} \left(\frac{1}{2}\int \phi_t\, dr\right) dr + A^1_t(t,\theta,z)\int e^{-V(r)} dr + A^2(t,\theta,z) \\
X^1 &= \frac{1}{2}\int \phi\, dr + A^1(t,\theta,z) \\
X^2 &= -\int e^{-U(r)} \left(\frac{1}{2}\int \phi_\theta\, dr\right) dr - A^1_\theta(t,\theta,z)\int e^{-U(r)} dr + A^3(t,\theta,z) \\
X^3 &= -\int e^{-W(r)} \left(\frac{1}{2}\int \phi_z\, dr\right) dr - A^1_z(t,\theta,z)\int e^{-W(r)} dr + A^4(t,\theta,z)
\end{aligned} \qquad (17)$$

where $A^1(t,\theta,z), A^2(t,\theta,z), A^3(t,\theta,z)$ and $A^4(t,\theta,z)$ are functions of integration. In order to determine $A^1(t,\theta,z), A^2(t,\theta,z), A^3(t,\theta,z)$ and $A^4(t,\theta,z)$ we need to integrate the remaining six equations. To avoid details, here we will present only results, when the above space-time (5) admits proper conformal vector fields. It follows after some tedious and lengthy calculations that there exists one case when the above space-time (5) admits proper conformal Killing vector field which are:

**Case (1)**

In this case the space-time (5) becomes

$$ds^2 = dr^2 + M^2(r)(-e^{-2d_7 N(r)} dt^2 + e^{-2d_{11} N(r)} d\theta^2 + e^{-2d_{14} N(r)} dz^2), \qquad (18)$$

where $M(r) = \frac{1}{2}\int \phi(r)\, dr + d_9$ and $N(r) = \int \frac{1}{M(r)}\, dr$. The conformal vector fields in this case are

$$\begin{aligned}
X^0 &= d_7 t + d_8 \\
X^1 &= M(r) \\
X^2 &= d_{11}\theta + d_{12} \\
X^3 &= d_{14} z + d_{15}
\end{aligned} \qquad (19)$$

where $d_7, d_8, d_9, d_{11}, d_{12}, d_{14}, d_{15} \in R (d_7 \neq 0, d_{11} \neq 0, d_{14} \neq 0, d_7 \neq d_{11}, d_7 \neq d_{14}, d_{11} \neq d_{14})$ and conformal factor is $\phi(r) = 2\frac{dM}{dr}$. The above space-time (18) admits four independent conformal vector fields in which one is proper conformal and three are independent Killing vector fields. The proper conformal vector field after subtracting Killing vector fields is

$$X = (d_7 t, M(r), d_{11}\theta, d_{14}z). \qquad (20)$$

Now consider $d_{11} = d_{14}$ and $d_7 \neq d_{11}$ and the space-time (18) becomes

$$ds^2 = dr^2 + M^2(r)(-e^{-2d_7 N(r)}dt^2 + e^{-2d_{11} N(r)}(d\theta^2 + dz^2)), \qquad (21)$$

The above space-time (18) admits five independent conformal vector fields in which four independent Killing vector fields which are $\frac{\partial}{\partial t}$, $\frac{\partial}{\partial \theta}$, $\frac{\partial}{\partial z}$ and $z\frac{\partial}{\partial \theta} - \theta\frac{\partial}{\partial z}$ and one proper conformal vector field which is given in equation (20). The cases when $d_7 = d_{11}, d_7 \neq d_{14}$ and $d_7 = d_{14}, d_{11} \neq d_{14}$ are exactly the same.